\documentclass{article}

\usepackage{latexsym}

\begin{document}

\begin{titlepage}
\baselineskip .15in

\begin{center}
{\bf

\vskip 1cm {\Large Conformal Collineations and Ricci Inheritance
Symmetry in String Cloud and String Fluids}

}\vskip .6in

{\sc U\u{g}ur Camci  }${}^{\dagger}$\\[1em]

{\em Department of Mathematics, Faculty of Arts and Sciences,

\d{C}anakkale Onsekiz Mart University, 17100 \d{C}anakkale,
Turkey}

\end{center}

\vskip 5cm

\begin{abstract}
Conformal collineations (a generalization of conformal motion) and
Ricci inheritance collineations, defined by $\pounds _\xi
R_{ab}=2\alpha R_{ab}$, for string cloud and string fluids in
general relativity are studied. By investigating the kinematical
and dynamical properties of such fluids and using the field
equations, some recent studies on the restrictions imposed by
conformal collineations are extended, and new results are found.
\end{abstract}

\vskip 1cm
\begin{center}
September, 2001
\end{center}
\vfill \baselineskip .2in $\dagger$~e-mail : ucamci@comu.edu.tr
\end{titlepage}

\normalsize

\baselineskip 15pt

\section{Introduction}
\label{int} \vspace*{-0.5pt} \noindent Let $M$ denotes a smooth
$n$-dimensional semi-Riemannian manifold with a non-degenerate
metric $g$ of arbitrary signature. Throughout this paper, all
geometric objects and their associated structures on $M$ will be
assumed smooth.

     A space-time on $M$ admits a one-parameter group of conformal motions
(Conf M) generated by a vector field $\xi$ if
\begin{equation}
\pounds _\xi g_{ab}=2\psi g_{ab}, \label{confm}
\end{equation}
where $\pounds _\xi$ signifies the Lie derivative operator along
$\xi$, and $\psi = \psi (x^a)$ is the conformal factor. The vector
field $\xi$ is, then, called a conformal Killing vector (CKV)
field which is a homothetic vector (HV) if $\psi_{,a} = 0$ and
Killing vector (KV) if $\psi = 0$. A special conformal Killing
vector (SCKV) field is restricted to $\psi_{;ab} = 0$. Here, (;)
and (,) denote the covariant and ordinary derivatives,
respectively. A Conf M defined by Eq. (\ref{confm}) satisfies
\begin{equation}
\pounds _\xi \Gamma _{bc}^a = \delta _b^a\psi _{;c} + \delta_c^a
\psi _{;b}-g_{bc} g^{ad} \psi _{;d}, \label{affinec}
\end{equation}
but the converse need not be true. A vector field $\xi$ satisfying
(\ref{affinec}) is called an affine conformal vector (ACV), which
is equivalent to \cite{duggal,duggal1}
\begin{equation}
\pounds _\xi g_{ab}=2\psi g_{ab} + H_{ab}, \quad H_{[a b]} = H_{a
b ; c} = 0, \label{confc}
\end{equation}
where $H_{ab}$ is a symmetric, covariant constant (and, therefore,
Killing) tensor associated with $\xi$. When $\psi_{;ab} = 0$, an
ACV is called special affine conformal vector (SACV) field. A
vector field $\xi$ satisfying (\ref{affinec}) or equivalently
(\ref{confc}) is  also called {\it conformal collineation} (Conf
C) \cite{duggal2}. A Conf C is reduced to CKV if $M$ is an
irreducible Riemannian manifold or a compact orientable Riemannian
manifold without boundary or a space of constant curvature. For a
Conf C defined by (\ref{affinec}), equivalently (\ref{confc}), the
following holds \cite{sharma}
\begin{eqnarray}
& &\pounds _\xi g^{ab} = - 2\psi g^{ab} - H^{ab}, \\& & \pounds
_\xi R_{ab} = -(n - 2)\psi _{;ab} -g_{ab}\Box \psi, \label{ric1}
\\& & \pounds _\xi R^{a}_{\,\,b} = -(n - 2)g^{ac} \psi_{;cb} -
\delta^a_b \Box \psi - 2 \psi R^{a}_{\,\,b} - H^{a}_{\,\,c}
R^{c}_{\,\,b}, \label{ric2} \\& & \pounds _\xi R = -2(n - 1) \Box
\psi - 2\psi R - R', \label{r1}
\end{eqnarray}
where $\Box $ is the Laplacian operator defined by $\Box \psi
=g^{ab}\psi _{;ab}$, and $R' \equiv H^{ab} R_{ab}$.

    Recently, Duggal introduced a new symmetry called {\it Ricci
inheritance collineation} (RIC) defined by \cite{duggal3}
\begin{equation}\label{ric}
\pounds _\xi R_{ab} = 2 \alpha R_{ab}
\end{equation}
where $\alpha = \alpha(x^a)$ is a scalar function. In particular,
RIC is reduced to {\it Ricci collineation} (RC) when $\alpha = 0$.
Then, using (\ref{confc}) and (\ref{ric}), we get the following
identities
\begin{eqnarray}
& & \pounds _\xi R^{a}_{\,\,b} = 2 (\alpha - \psi) R^a_{\,\,b} -
H^{a}_{\,\,c} R^{c}_{\,\,b}, \label{ric3}
\\& & \pounds _\xi R =  2 (\alpha - \psi ) R - R' . \label{r2}
\end{eqnarray}
It follows from (\ref{r1}) and (\ref{r2}) that
\begin{equation}
\Box \psi + \frac{\alpha}{n-1} R = 0 . \label{rson}
\end{equation}
Thus, equating (\ref{ric1}) and (\ref{ric}), using this result we
have
\begin{equation}
\psi_{;ab} = \frac{\alpha}{n-2} \left[ \frac{R}{n-1} g_{ab} - 2
R_{ab} \right] . \label{ricson}
\end{equation}

    The study of The CKVs and ACVs in a class of fluid space-times has recently
attracted some interest. Herrera {\it et al.}\cite{herrera}
studied CKVs and anisotropic fluids; Duggal and
Sharma\cite{duggal} extended this work to the more general case of
special ACV $\xi^a$. Maartens {\it et al.}\cite{mmt} investigated
the CKVs in anisotropic fluids, in which they are particularly
concerned with SCKVs. Mason and Maartens\cite{mm} considered
kinematics and dynamics of conformal collineations with the
general class of anisotropic fluids and no energy flux. Coley and
Tupper\cite{ct} discussed space-times admitting SCKV and symmetry
inheritance. Yavuz and Y{\i}lmaz\cite{yy1}, and Y{\i}lmaz {\it et
al.}\cite{y1999} have examined kinematic and dynamic properties of
string cloud and string fluids that admit a Conf M. Also, Baysal
{\it et al.}\cite{b} have been interested in Conf C for string
cloud, in the case of a SACVs. In this paper, we will extend the
previous obtained results\cite{y1999,b} to Conf C for string cloud
and string fluids, considering proper ACVs and RICs defined by
(\ref{ric}).

    The energy-momentum tensor for a cloud of string is
\begin{equation}
T_{ab} = \rho u_a u_b - \lambda \chi_a \chi_b \label{cloud}
\end{equation}
where $\rho $ is the rest energy for cloud of strings with
particles attached to them and $\lambda $ is string tensor density
and are related by
\begin{equation}
\rho = \rho_p + \lambda .
\end{equation}
Here $\rho_p$ is particle energy density. The unit timelike vector
$u^a$ describes the cloud four-velocity and the unit spacelike
vector $\chi^a$ represents a direction of anisotropy, i.e., the
string's directions \cite{let1}. For $u^a$ and $\chi^a$, we have
\begin{equation}
u^a u_a = - \chi^a \chi_a = -1, \quad \quad u^a \chi_a = 0.
\label{kosullar}
\end{equation}
The energy-momentum tensor for a fluid of strings\cite{let2,let3}
is
\begin{equation}
T_{ab}=\left( q + \rho_s \right)\left(u_a u_b - \chi_a \chi_b
\right)+ q g_{ab}, \label{fluid}
\end{equation}
where $\rho _s$ is string density and $q$ is {\it string tension}
and also {\it pressure}.

    The paper is organized as the following. In Sec. 2\ref{KM},
some kinematical results are summarized in the case of Conf C. In
Sec. 3\ref{dynamic}, using the field equations, dynamical
properties and equations of state for string cloud and string
fluid are derived by imposing a proper ACV and RIC. In final
Section, the obtained results are discussed.

\section{Some Kinematical Results}
\label{KM} \noindent The effect of ACV on any non-null unit vector
$X^a$ is given by
\begin{eqnarray}
\pounds_{\xi }X^{a} &=& - \left(\psi + \frac{\epsilon}{2} H_{bc}
X^b X^c \right) X^{a} + Y^{a}, \label{vector1} \\ \pounds _{\xi
}X_{a} &=& \left(\psi - \frac{\epsilon}{2} H_{bc} X^b X^c \right)
X_{a} + H_{ab} X^b + Y_{a}, \label{vector2}
\end{eqnarray}
where $Y^{a}$ is some vector orthogonal to $X^{a}$, i.e.,
$X^{a}Y_{a}=0$, $\epsilon = + 1$ if $X^a$ is spacelike and
$\epsilon = -1$ if $X^a$ is timelike. The proof of Eqs.
(\ref{vector1}) and (\ref{vector2}) which is a generalization of
that for the case of a CKV (see Ref. $7$) is obtained by Mason and
Maartens\cite{mm}. In general  $Y^a \neq 0$ : an explicit example
of a CKV ($H_{ab} = 0$) with $Y^a \neq 0$ in Robertson-Walker
spacetime is given in Ref. $7$.

    For the timelike unit four-velocity vector $u^a$ and  spacelike
unit vector $\chi^a$ of the string cloud and string fluids, we
have
\begin{eqnarray}
\pounds _\xi u^a &=& -\left( \psi - \frac{1}{2} H_{b c} u^b u^c
\right)u^a + v^a, \label{contra-u} \\ \pounds _\xi u_a &=& \left(
\psi + \frac{1}{2} H_{b c} u^b u^c \right)u_a + H_{a b} u^b + v_a
, \label{covar-u}
\end{eqnarray}
where $u^a u_a = -1, \quad u_a v^a = 0$ and
\begin{eqnarray}
\pounds _\xi \chi^a &=& - \left( \psi + \frac{1}{2} H_{b c} \chi^b
\chi^c \right)\chi^a + m^a , \label{contra-x} \\ \pounds _\xi
\chi_a &=& \left( \psi - \frac{1}{2} H_{b c} \chi^b \chi^c \right)
\chi_a + H_{ab} \chi^b + m_a , \label{covar-x}
\end{eqnarray}
where $\chi^a \chi_a = 1, \quad \chi_a m^a = 0$. Since $\chi_a u^a
= 0$ [see eq.(\ref{kosullar})] we get
\begin{equation}
\chi_a\pounds _\xi u^a + u^a \pounds _\xi \chi_a=0.
\label{lie-u-x}
\end{equation}
Substituting Eqs. (\ref{contra-u}) and (\ref{covar-x}) into
(\ref{lie-u-x}), yields
\begin{equation}
m_a u^a + v_a \chi^a = - H_{a b} u^a \chi^b . \label{constraint1}
\end{equation}
By (\ref{contra-u}) and (\ref{covar-u}), it follows that the
integral curves of an ACV $\xi$ are material curves iff $v^a = 0$.

\section{Dynamical Results and Equations of State}
\label{dynamic} \noindent Let $(V_4,g_{ab})$ be a spacetime of
general relativity. Then, we now consider how the Einstein's field
equations alter the purely kinematic results of the above section
for a Conf C. If $\xi^a$ is an ACV satisfying (\ref{confc}), then
\begin{eqnarray}
& & \pounds _{_\xi }R_{ab} = -2\psi _{;ab}-g_{ab}\Box \psi,
\label{lie-rab}
\\ & & \pounds _{_\xi }R = -2\psi R-6\Box \psi - R' , \label{lie-r}
\end{eqnarray}
where $R_{ab}$ is Ricci tensor and $R=g^{ab}R_{ab}$ is Ricci
scalar. Via Einstein's field equations
\begin{equation}\label{efe}
R_{ab}-\frac 12R g_{ab} = T_{ab}
\end{equation}
we find for the Lie derivative of an arbitrary energy-momentum
tensor $T_{ab}$
\begin{equation}
\pounds _{_\xi }T_{ab}=  -2\psi _{;ab} + 2 g_{ab}\Box \psi +
\frac{1}{2} \left( g_{ab} R' - H_{ab} R\right). \label{mcol}
\end{equation}

    By the use of (\ref{lie-rab}) in the known identity\cite{katzin}
\begin{equation}
\left( R^{ab} \xi_b \right)_{;a} \equiv \frac{1}{2} g^{ab}\pounds
_{_\xi }R_{ab} \label{k0}
\end{equation}
we obtain that
\begin{equation}
\left( R^{ab} \xi_b \right)_{;a} = \frac{1}{2} \left( \pounds
_{_\xi }R + 2 \psi R + R' \right). \label{k1}
\end{equation}
If $\xi$ is a ACV satisfying (\ref{confc}), then it follows by
(\ref{lie-r}) in the (\ref{k1}) that
\begin{equation}
\left( R^{ab} \xi_b \right)_{;a} = - 3 \Box \psi, \label{k2}
\end{equation}
and if $\xi$ is also RIC, then substituting $\Box \psi =
-\frac{\alpha}{3} R$ [see Eq. (\ref{rson})] in (\ref{k2}), gives
\begin{equation}
\left( R^{ab} \xi_b \right)_{;a} = \alpha R. \label{k3}
\end{equation}
By means of the Einstein field equations in the form
\begin{equation}
R^{ab} = T^{ab} - \frac{1}{2} T g^{ab}
\end{equation}
the condition (\ref{k2}) is expressible in the form of conserved
4-current
\begin{equation}
J^a_{\,\,;a} = 0, \label{k4}
\end{equation}
where the conserved vector $J^a$ is defined by
\begin{equation}
J^a = \left( T^{ab} - \frac{1}{2} T g^{ab} \right) \xi_b + 3
g^{ab} \psi_{,b}. \label{ja}
\end{equation}
Equation (\ref{k3}), or equivalently (\ref{k4}), is the basis for
generating new equations of state for various matter under
consideration.

\subsection{String Cloud}
\label{scloud} \noindent Take into account $T_{ab}$ to be of the
form (\ref{cloud}), with the aid of (\ref{contra-u}) for $\pounds
_{_\xi } u_a$ and (\ref{contra-x}) for $\pounds _{_\xi }
\chi_a$,\, a direct calculation yields
\begin{eqnarray}
\pounds _{_\xi }T_{ab}&=& \left[\pounds _\xi \rho + 2\psi \rho
\right] u_a u_b - \left[\pounds_{\xi }\lambda + 2\psi \lambda
\right] \chi_a \chi_b \nonumber \\ & & + H_{c d} \left( \rho u^c
u^d u_a u_b + \lambda \chi^c \chi^d \chi_a \chi_b \right) + H_{b
c} \left( \rho u^c u_a - \lambda \chi_a \chi^c \right) \nonumber
\\ & & + H_{a c} \left( \rho u_b u^c - \lambda \chi_b \chi^c \right) + 2
\rho u_{(a} v_{b)} -2\lambda \chi_{(a} m_{b)},
\end{eqnarray}
which, when substituted into (\ref{mcol}), gives
\begin{eqnarray}
-2\psi _{;ab}- \frac{1}{2} R H_{ab} + (2\Box \psi +
\frac{1}{2}R')g_{ab} &=& \left[\pounds _{_\xi }\rho + 2\psi \rho
\right]u_a u_b - \left[ \pounds _{_\xi }\lambda + 2\psi \lambda
\right] \chi_a \chi_b \nonumber
\\ & & +\rho \left[ H_{c d} u^d u_a u_b + H_{b c} u_a + H_{ac}
u_b \right] u^c \nonumber \\ & & +\lambda \left[ H_{c d} \chi^d
\chi_a \chi_b - H_{b c} \chi_a - H_{a c} \chi_b \right] \chi^c
\nonumber
\\ & & + 2\rho u_{(a}v_{b)}-2 \lambda \chi_{(a} m_{b)}.
\label{mcol-eq}
\end{eqnarray}
Now, we use $p_{ab}$ projection tensor that projects in the
directions that are perpendicular to both $\chi^a$ and $u^a$,
\begin{equation}
p_{ab} = g_{ab} + u_a u_b - \chi_a \chi_b.
\end{equation}
Some properties of this tensor are
\begin{equation}
p^{ab}u_{b} = p^{ab} \chi_{b} = 0
\end{equation}
\begin{equation}
p_{c}^{a} p_{b}^{c} = p_{b}^{a},\,\, p_{ab} = p_{ba}
\end{equation}
By contracting (\ref{mcol-eq}) with the tensors
$u^{a}u^{b},\chi^{a}\chi^{b}, p^{a b}, u^{a} \chi^{b}, u^{a}
p^{bc},\chi^{a} p^{bc},$ and $p^a_c p^b_d - \frac{1}{2}p^{ab}
p_{cd}$ the following equations for the string cloud are derived:
\begin{eqnarray}
\pounds_{_{\xi }}\rho + 2\psi \rho = -2\left(\Box \psi + \psi
_{;ab} u^a u^b \right)&+& \frac{1}{2}\rho_p H_{a b} u^a u^b -
\frac{1}{2} R' \label{mattereq1}
\\ \pounds _{_\xi }\lambda + 2\psi \lambda = -2\left(\Box \psi -
\psi _{;ab} \chi^a \chi^b \right)&+& \frac{1}{2} \rho_p H_{a b}
\chi^a \chi^b - \frac{1}{2} R' \label{mattereq2}
\\ \frac{1}{2} \left[ (\rho + \lambda)H_{ab} + 4 \psi_{;ab}
\right] p^{a b} &=& 4 \Box \psi + R',  \label{mattereq3}
\\ \rho_p \left[ \chi^a v_a + \frac{1}{2} H_{a b} u^a
\chi^b  \right] &=& 2 \psi _{;ab} u^a \chi^b, \label{mattereq4}
\\ \rho p^{b c} v_b + \frac{1}{2} \rho_p H_{a b} u^a
p^{b c} &=& 2 \psi _{;ab} u^a p^{bc}, \label{mattereq5} \\ \lambda
p^{bc} m_b -\frac{1}{2} \rho_p H_{a b} \chi^a p^{bc} &=& 2 \psi
_{;ab} \chi^a p^{bc}, \label{mattereq6} \\ \left[ (\rho + \lambda)
H_{ab} + 4 \psi_{;ab} \right] \left( p^a_c p^b_d -
\frac{1}{2}p^{ab} p_{cd} \right) &=& 0, \label{mattereq7}
\end{eqnarray}
where $R' = R^{cd} H_{cd}$, that is,
\begin{equation}
R' = \frac{1}{2} \left[ (\rho + \lambda) p^{cd} + \rho_p (u^c u^d
+ \chi^c \chi^d) \right] H_{cd}. \label{r'1}
\end{equation}
Equations (\ref{mattereq1})-(\ref{mattereq7}) are valid for any
ACV $\xi ^a$. Using (\ref{mattereq3}) in (\ref{mattereq7}) we have
\begin{equation}
\left[ (\rho + \lambda)H_{ab} + 4 \psi_{;ab} \right] p^a_c p^b_d =
( 4 \Box \psi + R') p_{c d},
\end{equation}
which gives Eq. (\ref{mattereq3}) by contracting with $g^{cd}$.
Therefore, Eqs. (\ref{mattereq3}) and (\ref{mattereq7}) are
equivalent.

    Now, using the string cloud energy-momentum tensor
(\ref{cloud}), it follows from Einstein's field equations
(\ref{efe}) that
\begin{equation}
R_{a b} u^a u^b = \frac{1}{2} \rho_p = R_{a b} \chi^a \chi^b,
\quad R_{a b} p^{a b} = \rho + \lambda. \label{contricci}
\end{equation}
Then, Eqs. (\ref{rson}), (\ref{ricson}), and (\ref{contricci})
give
\begin{equation}
\psi_{;ab} u^a u^b = \frac{\alpha}{3} (\lambda - 2 \rho), \,\,
\psi_{;ab} \chi^a \chi^b = \frac{\alpha}{3} (2 \lambda - \rho),
\,\, \psi_{;ab} p^{a b} = - \frac{2\alpha}{3} (\rho + \lambda),
\label{psiab1}
\end{equation}
\begin{equation}
\psi_{;ab} u^a \chi^b = 0, \,\, \psi_{;ab} u^a p^{b c} = 0,\,\,
\psi_{;ab} \chi^a  p^{b c} = 0. \label{psiab2}
\end{equation}
Thus, substituting $\Box \psi = -\frac{\alpha}{3} (\rho +
\lambda)$ [see Eq. (\ref{rson})] and (\ref{psiab1}) into Eqs.
(\ref{mattereq1}), (\ref{mattereq2}) and (\ref{mattereq3}), yield
\begin{eqnarray}
\pounds_{_{\xi }}\rho  + 2 (\psi - \alpha) \rho &=& -\frac{1}{4}
H_{ab} \left[ (\rho + \lambda) p^{ab} - 2 \rho_p u^a u^b \right],
\label{lie-ro}
\\ \pounds_{_{\xi }}\lambda + 2 (\psi - \alpha) \lambda &=& -\frac{1}{4}
H_{ab} \left[ (\rho + \lambda) p^{ab} - 2 \rho_p \chi^a \chi^b
\right] \label{lie-lamda}, \\ R' &=& \frac{1}{2} (\rho + \lambda)
H_{ab} p^{ab}. \label{r'2}
\end{eqnarray}
Using (\ref{r'2}) in (\ref{r'1}), we get
\begin{equation}
\rho_p \left( u^a u^b + \chi^a \chi^b \right) H_{ab} = 0.
\label{mattereq3-2}
\end{equation}
Furthermore, substituting (\ref{psiab2}) into equations
(\ref{mattereq4})-(\ref{mattereq6}), it follows that
\begin{eqnarray}
\rho_p \left[ \chi^a v_a + \frac{1}{2} H_{a b} u^a \chi^b  \right]
&=& 0, \label{mattereq4-2}
\\ \rho p^{b c} v_b + \frac{1}{2} \rho_p H_{a b} u^a
p^{b c} &=& 0, \label{mattereq5-2} \\ \lambda p^{bc} m_b
-\frac{1}{2} \rho_p H_{a b} \chi^a p^{bc} &=& 0.
\label{mattereq6-2}
\end{eqnarray}

   For a cloud of strings with $T_{ab}$ given in (\ref{cloud}), using
$R = (\rho + \lambda)$, we find from (\ref{k3}) that
\begin{equation}
2 \alpha (\rho + \lambda ) = \left( \left[ (\rho + \alpha) p^{ab}
+ \rho_p (u^a u^b + \chi^a \chi^b)\right] \xi_b \right)_{;a}.
\label{eq-state}
\end{equation}
Then, following three cases are taken (where we use the relation
$\xi^a_{;a} = 4\psi + \frac{1}{2} H^a_a$ obtained from
(\ref{confc})) :

Case (i) : If $\xi^a$ is orthogonal to $u^a$, i.e. $u^a \xi_a =
0$, thus, Eq. (\ref{eq-state}) provides
\begin{equation}
2 \alpha (\rho + \lambda) -4 \psi \rho_p = \pounds_{_{\xi }}\rho -
\pounds_{_{\xi }}\lambda + \frac{1}{2}\rho_p H^a_a \label{eq-s1}
\end{equation}
Then, substituting (\ref{lie-ro}) and (\ref{lie-lamda}) into this
equation, after some algebra, yields
\begin{eqnarray}
2 \alpha \lambda =  \rho_p \left[ \psi + \frac{1}{4}H^a_a +
\frac{1}{4} H_{ab}(u^a u^b - \chi^a \chi^b) \right] \nonumber
\end{eqnarray}
which, using (\ref{mattereq3-2}), becomes
\begin{equation}
8 \alpha \lambda =  \rho_p \left[ 4 \psi + H^a_a - 2 H_{ab}\chi^a
\chi^b \right]. \label{eqst-i}
\end{equation}

Case (ii) : If $\xi^a$ is parallel to $u^a$, i.e. $\xi^a = \xi
u^a$, then since $u^a \chi_a = 0$ and $p^{ab} u_a = 0$,  we obtain
from equation (\ref{eq-state}) that
\begin{equation}
2 \alpha (\rho + \lambda) + 4 \psi \rho_p = \pounds_{_{\xi
}}\lambda - \pounds_{_{\xi }}\rho - \frac{1}{2}\rho_p H^a_a
\label{eq-s2}
\end{equation}
which, when (\ref{lie-ro}) and (\ref{lie-lamda}) substituted into
(\ref{eq-s2}), and using (\ref{mattereq3-2}), gives
\begin{equation}
8 \alpha \rho = -\rho_p \left[ 4 \psi + H^a_a + 2 H_{ab}u^a u^b
\right]. \label{eqst-ii}
\end{equation}

Case (iii) : If $\xi^a$ is orthogonal to both $u^a$ and $\chi^a$,
then $p^{ab} \xi_b = \xi^a$, and Eq. (\ref{eq-state}) becomes
\begin{equation}
2 (\alpha -2\psi) (\rho + \lambda) = \pounds_{_{\xi }}\rho +
\pounds_{_{\xi }}\lambda + \frac{1}{2}(\rho + \lambda) H^a_a
\label{eq-s3}
\end{equation}
Proceeding as before we find
\begin{equation}
(\rho + \lambda)\left[ 4 \psi + H^a_a - H_{ab} p^{ab} \right] = 0.
\label{eqst-iii}
\end{equation}

        Let us consider the following possibilities according to whether
$\rho_p$ vanishes or not.

    Case (A) : $\rho_p = 0$, i.e. $\rho = \lambda$ (geometric string).

In this case, for $\rho \neq 0 \neq \lambda$, we have from Eqs.
(\ref{lie-ro})-(\ref{eq-state}) that
\begin{eqnarray}
& & \pounds_{_{\xi }}\rho  + 2 ( \psi - \alpha ) = - \frac{1}{2}
H_{ab} p^{ab} \rho, \label{a1} \\& & v^a = (\chi^b v_b) \chi^a,
\label{a2} \\& & m^a = - (u^b m_b) u^a, \label{a3}
\\ & & 2 \alpha \rho = \left( \rho p^{ab} \xi_b \right)_{;a}.
\label{a4}
\end{eqnarray}
When $\xi^a \perp u^a = 0$, i.e. $\xi^a u_a$, or $\xi^a \parallel
u^a$, i.e., $\xi^a = \xi u^a$, we find from the above equations
that $\alpha = 0$ and $\psi_{;ab} = 0$ so that RIC is reduced to
RC, that is, string cloud does not admit RIC, and $\xi^a$ must be
a SACV field. If $\xi^a$ is orthogonal to both $u^a$ and $\chi^a$,
then from (\ref{a1}) and (\ref{a4}), we obtain
\begin{eqnarray}
& & \pounds_{_{\xi }}\rho + 2 \left( \frac{1}{2}H_{ab} p^{ab} -
\frac{1}{4} H^a_a  - \alpha \right) \rho  = 0, \label{a5} \\& &
\psi = \frac{1}{4} \left( H_{ab} p^{ab} - H^a_a \right),
\label{a6}
\end{eqnarray}
which, assuming $H_{ab} u^b = \mu u_a$, i.e. $u^a$ is an
eigenvector of $H_{ab}$ and $H_{ab}\chi^b = \mu \chi_a$, i.e.
$\chi^a$ is an eigenvector of $H_{ab}$, yield
\begin{eqnarray}
& & \pounds_{_{\xi }}\rho + 2 \left( \frac{1}{4} H^a_a - \mu -
\alpha \right) \rho  = 0, \label{a7} \\& & \mu = - 2 \psi.
\label{a8}
\end{eqnarray}
where $\mu$ is an eigenvalue of $H_{ab}$.

    Case (B) : $\rho_p \neq 0$.

In this case, it  is assumed that
\begin{equation}
\rho \neq 0  \ne \lambda, \label{b1}
\end{equation}
and
\begin{equation}
H_{ab} u^b = \mu u_a , \label{b2}
\end{equation}
then from Eqs. (\ref{constraint1}) and
(\ref{mattereq3-2})-(\ref{mattereq6-2}), we have
\begin{equation}
H_{ab} \chi^b = \mu \chi_a \label{b3}
\end{equation}
and
\begin{equation}
v^a = 0 = m^a. \label{b4}
\end{equation}

    In this case, there are also three subcases :

Subcase (B.i) : $\xi^a \perp u^a$, i.e. $\xi^a u_a = 0$. In this
subcase, using the above obtained results in (\ref{lie-ro}),
(\ref{lie-lamda}), and (\ref{eqst-i}), we find
\begin{eqnarray}
\pounds_{_{\xi }}\rho  + 2 \left( \psi - \alpha - \frac{1}{8}
H^a_a \right) \rho & = & \left( \mu - \frac{1}{4} H^a_a \right)
\lambda, \label{lie-ro2}
\\ \pounds_{_{\xi }}\lambda + 2 \left( \psi - \alpha - \frac{1}{8}
H^a_a \right) \lambda & = &  \left( \mu - \frac{1}{4} H^a_a
\right) \rho,  \label{lie-lamda2}, \\ \alpha & = & \frac{\rho_p}{8
\lambda} \left[ 4 \psi -2 \mu + H^a_a \right]. \label{alfa1}
\end{eqnarray}

Subcase (B.ii) :$\xi^a \parallel u^a$, i.e., $\xi^a = \xi u^a$. In
this subcase, it is seen that $\pounds_{_{\xi }}\rho$ and
$\pounds_{_{\xi }}\lambda$ are same as in subcase (B.i). However,
from (\ref{eqst-ii}),  $\alpha$ is
\begin{equation}
\alpha = -\frac{\rho_p}{8 \rho} \left[ 4 \psi -2 \mu + H^a_a
\right]. \label{alfa2}
\end{equation}

Subcase (B.iii) : $\xi^a \perp u^a$ and $\xi^a \perp \chi^a$. For
this subcase, it follows from (\ref{lie-ro}), (\ref{lie-lamda}),
and (\ref{eqst-iii}) that
\begin{eqnarray}
\pounds_{_{\xi }}\rho + 2 \left( \frac{1}{8} H^a_a - \frac{1}{2}
\mu - \alpha \right) \rho &=& \left( \mu - \frac{1}{4} H^a_a
\right) \lambda, \label{lie-ro3}
\\ \pounds_{_{\xi }}\lambda + 2 \left( \frac{1}{8} H^a_a -
\frac{1}{2}\mu - \alpha \right) \lambda &=& \left( \mu -
\frac{1}{4} H^a_a \right)\rho, \label{lie-lamda3}, \\ \mu &=& -2
\psi. \label{psi3}
\end{eqnarray}

\subsection{String Fluid}
\label{sfluid} \noindent In this subsection, following exactly as
in the case of string cloud, taking the Lie derivative of $T_{ab}$
given by (\ref{fluid}) for string fluid and contracting in turn by
the tensors $u^{a}u^{b}$, $\chi^{a}\chi^{b}$, $p^{a b}$, $u^{a}
\chi^{b}$, $u^{a} p^{bc}$,$\chi^{a} p^{bc}$,and $p^a_c p^b_d -
\frac{1}{2}p^{ab} p_{cd}$, after a number of calculations, we get
\begin{eqnarray}
& & \pounds_{_{\xi }}\rho_s + 2\psi \rho_s = -2 \Box \psi - \left[
2 \psi _{;ab} - q H_{a b} \right] u^a u^b - \frac{1}{2} R'
\label{meq1}
\\& & \pounds _{_\xi }\rho_s + 2\psi \rho_s = -2 \Box \psi + \left[
2 \psi _{;ab} - q H_{a b} \right] \chi^a \chi^b - \frac{1}{2} R'
\label{meq2} \\& & \pounds _{_\xi } q + 2\psi q = 2 \Box \psi -
\frac{1}{2} \left[ 2\psi _{;ab} + \rho_s H_{ab} \right] p^{ab} +
\frac{1}{2} R', \label{meq3}
\\& & \left[ 2 \psi_{;ab} - q H_{ab} \right] u^a \chi^b = 0,
\label{meq4}
\\& & (\rho_s + q) p^{b c} v_b  = \left[ 2 \psi_{;ab} - q H_{a b} \right] u^a
p^{b c}, \label{meq5} \\& & (\rho_s + q) p^{bc} m_b = \left[ 2
\psi_{;ab} - q H_{a b} \right] \chi^a p^{bc}, \label{meq6}
\\& & \left[ 2 \psi_{;ab} + \rho_s H_{ab} \right] \left(
p^a_c p^b_d - \frac{1}{2}p^{ab} p_{cd} \right) = 0, \label{meq7}
\end{eqnarray}
where $R' = R^{cd} H_{cd}$, that is,
\begin{equation}
R' = (\rho_s + q ) H_{cd} p^{cd} - q H^a_a.
\end{equation}
Equating Eqs. (\ref{meq1}) and (\ref{meq2}), we find
\begin{equation}
\left[ 2 \psi_{;ab} - q H_{a b} \right] \left( u^a u^b + \chi^a
\chi^b \right) = 0. \label{meq-12}
\end{equation}
Using (\ref{fluid}) in (\ref{efe}), the following relations are
derived
\begin{equation}
R_{ab} u^a u^b = q = - R_{ab} \chi^a \chi^b, \quad R_{ab} p^{ab} =
2 (\rho_s - q)
\end{equation}
which, from Eqs. (\ref{rson}) and (\ref{ricson}), yields
\begin{equation}
\psi_{;ab} u^a u^b = -\frac{\alpha}{3} (\rho_s + 2 q) =
-\psi_{;ab} \chi^a \chi^b, \quad \psi_{;ab} p^{ab} =
-\frac{2\alpha}{3} (2 \rho_s + q), \label{ps-ab1}
\end{equation}
\begin{equation}
\psi_{;ab} u^a \chi^b = 0, \quad \psi_{;ab} u^a p^{b c} = 0,\quad
\psi_{;ab} \chi^a  p^{b c} = 0. \label{ps-ab2}
\end{equation}
Then, substituting $\Box \psi = -\frac{2\alpha}{3} (\rho_s - q)$
[see Eq. (\ref{rson})], (\ref{ps-ab1}) and (\ref{ps-ab2}) into
Eqs. (\ref{meq1})-(\ref{meq-12}),  gives
\begin{eqnarray}
& & \pounds_{_{\xi }}\rho_s + 2(\psi - \alpha) \rho_s = -
\frac{1}{2} \rho_s H_{ab} p^{ab}, \label{meq1-2}
\\& & \pounds _{_\xi } q + 2(\psi - \alpha) q = \frac{1}{2} q
\left[ H_{ab} p^{ab} - H^a_a \right], \label{meq3-2}
\\& & q H_{ab} \left( u^a u^b + \chi^a \chi^b \right) = 0,
\label{meq12-2} \\& & q H_{ab} u^a \chi^b = 0, \label{meq4-2} \\&
& (\rho_s + q) p^{b c} v_b  = - q H_{a b} u^a p^{b c},
\label{meq5-2} \\& & (\rho_s + q) p^{bc} m_b = - q H_{a b} \chi^a
p^{bc}, \label{meq6-2} \\& & \rho_s H_{ab} \left( p^a_c p^b_d -
\frac{1}{2}p^{ab} p_{cd} \right) = 0. \label{meq7-2}
\end{eqnarray}

    For a fluid of strings with $T_{ab}$ given in (\ref{fluid}),
using $R = 2 ( \rho_s - q)$ in (\ref{k3}), we get
\begin{equation}
2 \alpha (\rho_s  - q) = \left( \left[ (\rho_s + q) p^{ab} - q
g^{ab} \right] \xi_b \right)_{;a}. \label{eq-state-s}
\end{equation}

Now, we consider the cases (i)-(iii) given in string cloud. In
both cases (i), $\xi^a u_a = 0$,  and (ii), $\xi^a = \xi u^a$,
Eqs. (\ref{meq1-2}), (\ref{meq3-2}), and (\ref{eq-state-s}) yield
\begin{equation}
4 (\alpha \rho_s + \psi q ) = - q H_{ab} p^{ab}.
\label{eqst-i-ii-s}
\end{equation}
In case (iii), $\xi^a u_a = 0 = \xi^a \chi_a$, from
(\ref{meq1-2}), (\ref{meq3-2}), and (\ref{eq-state-s}), we find
\begin{equation}
4 (\alpha q + \psi \rho_s) = \rho_s H_{ab} \left(u^a u^b - \chi^a
\chi^b \right). \label{eqst-iii-s}
\end{equation}
Thus, in the case of geometric string, that is,
\begin{equation}
\rho_p = 0, \quad and \quad q = 0, \quad i.e. \,\, \rho = \rho_s =
\lambda  \label{kabul-1}
\end{equation}
we obtain the same relations given in (\ref{a1})-(\ref{a4}). If we
assume
\begin{equation}
\quad q \neq 0 \neq \rho_s \label{kabul-2}
\end{equation}
and
\begin{equation}
 H_{ab} u^b = \mu \, u_a,
\end{equation}
then, it is found from (\ref{constraint1}),
(\ref{meq1-2})-(\ref{meq7-2}) that
\begin{eqnarray}
& & \pounds _{_\xi } \rho_s +  2 \left( \psi - \alpha -
\frac{1}{2}\mu + \frac{1}{4} H^a_a \right) \rho_s = 0, \label{sf1}
\\& & \pounds _{_\xi } q  + 2\left( \psi - \alpha + \frac{1}{2} \mu \right) q = 
0,
\label{sf2} \\& & H_{ab} \chi^b = \mu \chi^a,  \quad u^a m_a = -
\chi^a v_a, \label{sf3} \\& & v^a = (\chi^a v_b) \chi^a, \quad m^a
= -(u^b m_b) u^a. \label{sf4}
\end{eqnarray}
Therefore, using these results, we obtain from Eq.
(\ref{eqst-i-ii-s}) that
\begin{equation}
\alpha = \frac{q}{4 \rho_s} \left[ 4 \psi - 2 \mu + H^a_a \right],
\label{sf5}
\end{equation}
for cases (i) and (ii), and from Eq. (\ref{eqst-iii-s})
\begin{equation}
\alpha = - \left( \psi + \frac{1}{2} \mu \right) \frac{\rho_s}{q}.
\label{sf6}
\end{equation}
for case (iii). The derivations of (\ref{sf1})-(\ref{sf6})
required the reasonable assumptions contained in (\ref{kabul-2})
since, when $\rho_s + q = 0$, fluid of strings disappear.

\section{Discussions and Conclusions}
\noindent In this paper, we have used the Einstein field equations
to investigate the dynamic restrictions imposed by the proper ACV
and RIC, and derived equations of state for string cloud and
string fluids. In order to find physically meaningful equations of
state, we need a relation between $\alpha, \psi$ and $H^a_a$. For
this purpose, we define {\it symmetry inheritance} of an ACV to
mean that a kinematical or a  dynamical variable  ${\it F}$, say,
satisfies an equation of the form
\begin{equation}
\pounds _{_\xi } {\it F} + (k \psi + r \alpha + s H^a_a ){\it F} =
0, \label{symm-inh}
\end{equation}
where $k, r$ and $s$ are constants, possibly depending on the
tensorial character of ${\it F}$. Symmetry inheritance for a CKV
in the above sense has been considered in Refs. $9$ and $17$. It
is an appealing idea because it relates the symmetry with all
kinematical and dynamical variables making full use of the field
equations. If ${\it F}$ satisfies Eq. (\ref{symm-inh}), then the
symmetries of an ACV $\xi^a$ are inherited. If this condition does
not hold, then the symmetries are not inherited. Also, in both of
the cases of string cloud and string fluids, we conclude that in
order for the dynamical variables to inherit the symmetry of an
ACV, the value of the constant $r$ is $-2$. The constants $k$ and
$s$ behave differently for considered cases.

(a) \,In the case of string cloud, we have found Eqs.
(\ref{eqst-i}), (\ref{eqst-ii}) and (\ref{eqst-iii}) from the
restrictions (i), (ii), and (iii) on the equation of state,
imposed by the field equations. Then, in case (A), i.e. $\rho =
\lambda$ (geometric string), we have also found $\alpha = 0$ and
$\psi_{;ab} = 0$ when $\xi^a \perp u^a$ and $\xi^a = \xi u^a$, so
that there exist no RIC other then RC, that is, string cloud does
not admit RIC. When $\xi^a u_a = 0 = \xi^a \chi_a$, assuming $u^a$
and $\chi^a$ are eigenvectors of $H_{ab}$, it is obtained
(\ref{a5}) as inheritance equation. In this case, if we assume
that
\begin{equation}
\mu = \frac{1}{4} H^a_a, \label{inh-cond}
\end{equation}
then our symmetry inheritance equation (\ref{symm-inh}) are
reduced to the same definitions in Refs. $9$ and $17$. Also, in
case (A), it is found that $v^a$ and $m^a$ are spacelike and
timelike vector fields, respectively. In case (B), i.e. $\rho_p
\neq 0$, it is obtained that there are two important cases when
$u^a$ and $\chi^a$ are eigenvectors of $H_{ab}$, so that $v^a$ and
$m^a$ vanish by (\ref{b4}) (provided the physically reasonable
energy condition (\ref{b1}) is satisfied). When $\alpha \neq 0$
(i.e., proper RIC), Eqs. (\ref{eqst-i}), (\ref{eqst-ii}) and
(\ref{eqst-iii}) provide physically meaningful equations of state
for given $\alpha, \psi$, and $H^a_a$.

In subcase (B.i), under the assumption (\ref{inh-cond}), from Eqs.
(\ref{lie-ro2}), (\ref{lie-lamda2}) and (\ref{alfa1}),  we obtain
that
\begin{eqnarray}
& & \pounds _{_\xi } \rho +  2 \left( \psi - \alpha - \frac{1}{8}
H^a_a \right) \rho = 0, \label{s1} \\& & \pounds _{_\xi } \lambda
+ 2 \left( \psi - \alpha - \frac{1}{8} H^a_a \right) \lambda = 0,
\label{s2} \\& & \alpha = \frac{\rho_p}{2\lambda} \left[ \psi +
\frac{1}{8} H^a_a \right]. \label{s3}
\end{eqnarray}
The equation (\ref{s3}) yields the following equation of state for
$\psi \neq - \frac{1}{4} H^a_a$
\begin{equation}
\rho = (1 + w) \lambda \label{takabayashi}
\end{equation}
which coincide with Takabayashi string, where $w$ is given by
\begin{equation}
w = \frac{2\alpha}{\psi + \frac{1}{8} H^a_a}.
\end{equation}
In subcase (B.ii), $\pounds _{_\xi } \rho$ and $\pounds _{_\xi }
\lambda$ are same as in subcase (B.i), but assuming the condition
(\ref{inh-cond}), $\alpha$ is
\begin{equation}
\alpha = - \frac{\rho_p}{2 \rho} \left[ \psi + \frac{1}{8} H^a_a
\right] \label{s4}
\end{equation}
giving equation of state (\ref{takabayashi}) with
\begin{equation}
w = - \frac{2\alpha}{2 \alpha + \psi + \frac{1}{8} H^a_a}.
\end{equation}
In subcase (B.iii), using the condition (\ref{inh-cond}) in
(\ref{lie-ro3}) and (\ref{lie-lamda3}), we have
\begin{eqnarray}
& & \pounds _{_\xi } \rho - 2 \alpha \rho = 0, \label{s5}
\\& & \pounds _{_\xi } \lambda  - 2\alpha \lambda = 0. \label{s6}
\end{eqnarray}

(b) In the case of string fluids, when $\rho_p = 0$ and $q = 0$,
i.e. geometric (Nambu) string, same relations in string cloud are
found. When $q \neq 0 \neq \rho_s$, it is found that $v^a$ and
$m^a$ are spacelike and timelike vector fields,respectively, and
Eqs. (\ref{sf1}) and (\ref{sf2}) are inheritance equations with $k
= 2, \, r = -2$ and $s = \frac{1}{8}$ if the condition
(\ref{inh-cond}) holds. Then, assuming the condition
(\ref{inh-cond}), it follows from (\ref{sf1}) and (\ref{sf2}) that
\begin{eqnarray}
& & \pounds _{_\xi } \rho_s +  2 \left( \psi - \alpha +
\frac{1}{8} H^a_a \right) \rho_s = 0, \label{s7}
\\& & \pounds _{_\xi } q  + 2\left( \psi - \alpha + \frac{1}{8} H^a_a \right) q 
=
0. \label{s8}
\end{eqnarray}
Hence, in both cases (i) and (ii), we have found the following
equation of state
\begin{equation}
q = \gamma \rho_s \label{sf-eq-of-state}
\end{equation}
where we have defined $\gamma$ as
\begin{equation}
\gamma = \frac{\alpha}{\psi + \frac{1}{4} H^a_a}. \label{g1}
\end{equation}
with $\psi \neq - \frac{1}{4} H^a_a$. Then, using (\ref{g1}), Eqs.
(\ref{s7}) and (\ref{s8}) become
\begin{eqnarray}
& & \pounds _{_\xi } \rho_s +  2 \left( \frac{1-\gamma}{\gamma}
\right) \alpha \rho_s = 0, \label{s9}
\\& & \pounds _{_\xi } q  + 2 \left( \frac{1-\gamma}{\gamma}
\right) \alpha q = 0,  \label{s10}
\end{eqnarray}
where $\gamma \neq 0$. For case (iii), $\gamma$ is defined as
\begin{equation}
\gamma = - \frac{\psi + \frac{1}{4} H^a_a}{\alpha} \label{g2}
\end{equation}
giving (\ref{sf-eq-of-state}). Therefore, substituting (\ref{g2})
into Eqs. (\ref{s7}) and (\ref{s8}), yields
\begin{eqnarray}
& & \pounds _{_\xi } \rho_s -  2 \left( 1 + \gamma \right) \alpha
\rho_s = 0, \label{s11}
\\& & \pounds _{_\xi } q  - 2 \left( 1 + \gamma \right) \alpha q = 0. \label
{s12}
\end{eqnarray}
Finally, it is easily seen that if $\gamma = 1$, i.e. $q =
\rho_s$, in cases (i) and (ii), or if $\gamma = -1$, i.e. $q +
\rho_s = 0$, in case (iii), then $\pounds _{_\xi } \rho_s = 0$ and
$\pounds _{_\xi } q = 0$.

\end{document}